\documentstyle[12pt,epsfig]{article}
\renewcommand{\theequation}{\arabic{section}.\arabic{equation}}

\makeatletter
\def\slash#1{{\mathpalette\c@ncel{#1}}} 
\makeatother

\newcommand\beq{\begin{eqnarray}}
\newcommand\eeq{\end{eqnarray}}

\newcommand\la{\langle}
\newcommand\ra{\rangle}

\def\Sslash{\rlap/{\mkern-1mu S}}
\def\pslash{\rlap/{\mkern-1mu p}}

\def\kslash{\slash{\mkern-1mu k}}

\def\Sslash{\slash{\mkern-1mu S}}

\def\xhat{\widehat{x}}
\def\zhat{\widehat{z}}

\def\pvec{\vec{p}}

\def\shat{\widehat{s}}
\def\dhat{\widehat{d}}

\def\uhat{\widehat{u}}

\begin{document}
\hspace{11.5cm}
\vbox{hep-ph/0302061}
\vspace*{7mm}
\begin{center}
{\Large \bf 
Double Spin Asymmetries for Large-$p_T$ Hadron Production 
in Semi-Inclusive DIS\\}
\vspace{1cm}
 {\sc Yuji~Koike and J. Nagashima}
\\
\vspace*{0.1cm}{\it Department of Physics, Niigata University,
Ikarashi, Niigata 950--2181, Japan}
\\[1cm]
\parbox[t]{\textwidth}{{\bf Abstract:} 
We study the twist-2 double-spin asymmetries for the 2-jets and
large-$p_T$ hadron ($\pi$, $\Lambda$ etc)
production in semi-inclusive deep inelastic scattering
to $O(\alpha_s)$ in perturbative QCD.
After deriving the complete set of the polarized cross section
which is differential with respect to the transverse momentum,
we discuss characteristic features of the 
azymuthal spin asymmetries,
using existing parton densities and fragmentation functions
at COMPASS and EIC energies.
}

\end{center}

\newpage

\section{Introduction}
\setcounter{equation}{0}
High energy experiment with polarized beams and targets has opened a
new window for revealing QCD dynamics and hadron structures.
Ongoing RHIC-SPIN, HERMES and COMPASS experiments are 
going to provide us with a variety of
data disclosing spin distributions inside the nucleon.
The planed Electron Ion Collider (EIC) at BNL is expected to be 
a more powerful and sensitive tool for the QCD spin physics.

In this paper, we study the hadron production
in polarized semi-inclusive deep inelastic scattering (SIDIS)
off proton.
The cross section formula with integrated transverse
momentum of the final hadron, $p_T$, has been derived
in the leading order (LO)\,\cite{Ji94} 
and in the next-to-leading order (NLO)\,\cite{FGS}
and has been applied to predict spin asymmetries 
for polarized $\Lambda$ production in SIDIS\,\cite{FSV98}.  
We extend these studies to derive cross section formula 
for the complete set of the polarized processes
which is differential in $p_T$ and is suited, in particular, for the 
large-$p_T$ hadron production.  
This study should be useful to get further insight into the parton densities
and fragmentation functions from HERMES, 
COMPASS and future EIC experiment.  It is also complementary
to studies on large-$p_T$ hadron production in $pp$ collisions 
at RHIC and HERA-$\vec{N}$\,\cite{FSV98pp,FSSV98,Boros}.  
As typical SIDIS processes, we study
pion (or 2-jets) and $\Lambda$ hyperon production.
In the leading twist-2 level,
the complete set of the processes is 
\beq
&{\rm (i)}&\quad e+p\rightarrow 
e'+\pi(p_T)+X\ {\rm or}\ \ e'+{\rm 2\ jets} +X,\nonumber\\
&{\rm (ii)}&\quad e+\vec{p}\rightarrow e'+ \vec{\Lambda}(p_T)+X,\nonumber\\
&{\rm (iii)}&
\quad e+ {p}^\uparrow\rightarrow e'+ {\Lambda}^\uparrow(p_T) +X,\nonumber\\
&{\rm (iv)}&
\quad \vec{e}+ \vec{p}\rightarrow e'+ \pi(p_T)+X\ {\rm or}\ \ 
e'+{\rm 2\ jets} +X,
\nonumber\\
&{\rm (v)}&\quad \vec{e}+ p\rightarrow e'+ \vec{\Lambda}(p_T) +X,
\label{eq1.1}
\eeq
where we have used the notation $\vec{p}$ and $\vec{\Lambda}$ for
longitudinally polarized, and ${p}^\uparrow$ and ${\Lambda}^\uparrow$
for transversely polarized hadrons, and we restrict ourselves
to electron (or muon) scattering off proton.
In order that a final hadron carries large transverse momentum, another parton
has to be emitted 
in the opposite direction.  This is an $O(\alpha_s)$ effect
in perturbative QCD and
was investigated in \cite{Men,MOS92} for the unpolarized
SIDIS with electron and neutrino beams.  
Here we extend the analysis to the polarized cases shown above.
For small-$p_T$ production in SIDIS, 
the origin of $p_T$ may be ascribed to the nonperturbative
intrinsic transverse
momentum of partons confined in or fragementing into a hadron.
Derivation of the cross section formula based on this idea
was performed in \cite{MT96},
which is complementary to the present study.

As will be shown below, our cross section formula for (i)-(v)
applicable in the large-$p_T$ 
region diverges as $1/p_T^2$ as $p_T\to 0$,
which is a manifestation of soft and collinear divergence.  
For the $p_T$-integrated cross section, the soft divergence
is canceled by the soft divergence in the 
virtual correction diagrams and the collinear one is factorized
into the parton densities.   
Accordingly, $O(\alpha_s)$ cross section at $p_T\approx 0$ is to be
interpreted as ``distribution'' with respect to the variable $p_T$.   
In order to get a finite $p_T$-differential cross section, one needs to
resum the effect of soft gluon radiation
as was studied in \cite{MOS96,NSY00}.  Derivation of the 
resummed cross section for a low $p_T$ region is
beyond the scope of the present
study and will be presented elesewhere.

This paper is organized as follows.  In section 2 we present
the formalism to calculate the cross sections for (\ref{eq1.1})
following \cite{MOS92}. 
Section 3 presents the final analytic formula for the cross section.
In Section 4, we present a numerical estimate of the spin asymmetry
and discuss its characteristic features, 
using existing parton distribution and fragmentation functions 
in the literature.
Section 5 is devoted to summary and conclusion.

\section{Formalism}
\setcounter{equation}{0}

In this paper we are interested in the cross section for the process,
\beq
e(k)+A(p_A,S_A) \rightarrow e'(k') + B(p_B,S_B)+X.
\label{eq2.1}
\eeq
We define 5 Lorentz invariants for specifying the kinematics.  
The center of mass energy $S_{ep}$ for the initial
electron and the proton $A$ is
\beq
S_{ep}=(p_A + k)^2 \simeq 2p_A\cdot k,
\eeq
where we ignored the masses.
Conventional DIS variables
\beq
x_{bj}={Q^2\over 2p_A\cdot q}, \qquad Q^2 =-q^2=-(k-k')^2,
\eeq
are determined by observing the final electron.
For the description of kinematics of the final hadron $B$, we introduce
\beq
z_f={p_A\cdot p_B\over p_A\cdot q},
\eeq
and the ``transverse'' component of $q$ which is orthogonal to
$p_A$ and $p_B$,
\beq
q_t^\mu=q^\mu- {p_B\cdot q\over p_A\cdot p_B}p_A^\mu - 
{p_A\cdot q\over p_A\cdot p_B}p_B^\mu.
\eeq
$q_t$ is a space-like vector and we define its magnitude as
\beq
q_T = \sqrt{-q_t^2}.
\eeq

Differential cross section for the semi-inclusive production 
(\ref{eq2.1}) can be written as
\beq
d\sigma= {1\over 2S_{ep}} {d^3p_B\over(2\pi)^3 2p_B^0}
{d^3k'\over (2\pi)^3 2k'^0} {e^4\over q^4}
L_{\mu\nu}(k,k')W^{\mu\nu}(p_A,p_B,q),
\label{eq2.2}
\eeq
where  
we use the covariant normalization for each state and 
$1/2S_{ep}$ is the initial flux. 
$L_{\mu\nu}$ is the leptonic tensor defined as
\beq
L_{\mu\nu}(k,k')&=&{1\over 2}{\rm Tr}\left[\gamma_\mu(1\pm\gamma_5)\kslash
\gamma_\nu\kslash'\right]\nonumber\\
&=&2(k_\mu k'_\nu + k_\nu k'_\mu) -g_{\mu\nu}Q^2 \pm 2i\epsilon_{\mu\nu\lambda
\sigma}k^\lambda k'^\sigma. 
\label{eq2.3}
\eeq
with our convention for the anti-symmetric tensor $\epsilon_{0123}
=-\epsilon^{0123}=1$.  
The leading twist-2 contribution to
the hadronic tensor
$W^{\mu\nu}$ can be written as
\beq
W^{\mu\nu}(p_A,p_B,q)&=&\int_0^1\,{dx\over x}\int_0^1\,{dz\over z^2}
{\rm Tr}\left[M_A(x,p_A,S_A)\widehat{M}_B(z,p_B,S_B)H^{\mu\nu}(xp_A,p_B/z,q)
\right]\nonumber\\
& &\qquad\times2\pi\delta\left((xp_A+q-p_B/z)^2\right),
\label{eq2.4}
\eeq
where $M_A(x,p_A,S_A)$ and $\widehat{M}_B(z,p_B,S_B)$ are
the distribution and fragmentation functions for the hadrons $A$ and $B$,
respectively, and $H^{\mu\nu}(xp_A,p_B/z,q)$ is the corresponding
hard part.
${\rm Tr}$ indicates the trace over
relevant spinor or Lorentz indices.  
Since we
are interested in the twist-2 cross sections, 
$p_A$ and $p_B$ can be regarded as light-like.  
To define the complete set of 
distribution and fragmentation functions, we introduce another set of
light-like vectors $n$ ($n^2=0$) and $w$ ($w^2=0$)
for $p_A$ and $p_B$, respectively, by the relation
$p_A\cdot n =1$ and $p_B\cdot w =1$.
The complete set of the 
twist-2 quark distribution for the proton $A$ is defined 
as\,\cite{CS82,JJ92}
\beq
M_{Aij}^{q}(x,p_A,S_A)&=& \int {d\lambda\over 2\pi}\,e^{i\lambda x}
\la p_A S_A | \bar{\psi}_j(0)\psi_i(\lambda n) |p_A S_A\ra\nonumber\\
&=&{1\over 2}\pslash_A q(x) + {1\over 2}\lambda_A\gamma_5\pslash_A
\Delta q(x)+ {1\over 2}\gamma_5 \Sslash_{A\perp} \pslash_A \delta q(x)+\cdots,
\eeq
where the transverse spin vector $S_{A\perp}^\mu =(0,\vec{S}_{A\perp})$
satisfies $\vec{S}_{A\perp}\cdot \vec{p}_A=0$ and 
$\lambda_A^2 + \vec{S}_{A\perp}^2=1$ with $\lambda_A$ the longitudinal
component of the spin vector, and $+\cdots$ stands for twist-3 or higher.
$q(x)$, $\Delta q(x)$ and $\delta q(x)$ are, respectively, 
spin-average, longitudinally polarized and transversely 
polarized (transversity) quark distribution.  
$\delta q$ is chiral-odd and hence does not mix with gluon distributions. 
The gluon distrbition is also defined as\,\cite{CS82,Mano90,Ji92}
\beq
M_A^{g\alpha\beta}(x,p_A,S_A)&=& {2\over x^2}
\int {d\lambda\over 2\pi}\,e^{i\lambda x}
\la p_A S_A|{\rm tr}\left[
n_\rho G^{\rho\alpha}(0)n_\sigma G^{\sigma \beta}(\lambda n)\right]
|p_A S_A\ra\nonumber\\
&=& -{1\over 2x}G(x) g_{A\perp}^{\alpha\beta} -{1\over 2x}\Delta G(x)
i\epsilon^{\alpha\beta\rho\sigma}p_{A\rho}n_\sigma+\cdots,
\eeq
where $G^{\rho\alpha}$ is the gluon's field strength,
tr means the trace over the color index for $G^{\rho\alpha}$,
$g_{A\perp}^{\alpha\beta}=g^{\alpha\beta}-p_A^\alpha n^\beta - 
n^\alpha p_A^\beta$. 
$G(x)$ and $\Delta G(x)$ are, respectively,
spin-average and longitudinally polarized gluon distributions. 
Notice there is no transversely polarized gluon distribution. 
Similarly we define twist-2 quark and gluon fragmenation functions
for the spin-1/2 hadron $B$\,\cite{CS82,JJ93}:
\beq
\widehat{M}^q_{Bij}(z,p_B,S_B)&=&{1\over N_c}
\sum_X\int {d\lambda \over 2\pi}\,
e^{-i\lambda/z}\la 0|\psi_i(0)|BX\ra\la BX|\bar{\psi}_j(\lambda w)|0\ra
\nonumber\\
&=&{1\over z}\pslash_B \widehat{q}(z)
+{1\over z}\lambda_B\gamma_5 \pslash_B \Delta \widehat{q}(z)
+{1\over z}\gamma_5\Sslash_{B\perp} \pslash_B \delta \widehat{q}(z)
+\cdots,
\eeq
\beq
\widehat{M}_B^{g\alpha\beta}
(z,p_B,S_B)&=&{2z^2\over N_c^2-1}\sum_X\int {d\lambda \over 2\pi}\,
e^{-i\lambda/z}\la 0|{\rm tr}[w_\rho G^{\rho\alpha}(0)|BX\ra\la BX|w_\sigma 
G^{\sigma\beta}(\lambda w)] |0\ra\nonumber\\
&=& -\widehat{G}(z) g_{B\perp}^{\alpha\beta} -\Delta \widehat{G}(z)
i\epsilon^{\alpha\beta\rho\sigma}p_{B\rho}w_\sigma+\cdots,
\eeq
where the transverse spin vector $S_{B\perp}^\mu=(0,\vec{S}_{B\perp})$
in the quark distribution satisfies
$\vec{S}_{B\perp}\cdot \vec{p}_B=0$ 
and $\lambda_B^2 + \vec{S}_{B\perp}^2=1$, and
$g_{B\perp}^{\alpha\beta}=g^{\alpha\beta}-p_B^\alpha w^\beta - 
w^\alpha p_B^\beta$. 
Physical meaning of
each fragmentation function is in parallel with the distribution
functions defined above, and we specify them 
by putting ``$\ \ \widehat{}\ \ $'' on the corresponding distributions.

For the actual calculation of the cross section, 
it is convenient to
work in the {\it hadron frame}\,\cite{MOS92}. 
In this frame $q^\mu$ and $p_A^\mu$
take
\beq
q^\mu &=& (0,0,0,-Q),\\
p_A^\mu &=& \left( {Q\over 2x_{bj}},0,0,{Q\over 2x_{bj}}\right).
\eeq
Choosing the $xz$-plane as the hadron plane, $p_B$ can be written as
\beq
p_B^\mu = {z_f Q \over 2}\left( 1 + {q_T^2\over Q^2},{2 q_T\over Q},0,
{q_T^2\over Q^2}-1\right).
\label{eq2.p_B}
\eeq
In order to write the lepton momentum in the hadron frame, we need
to introduce 
the azymuthal angle $\phi$
between the hadron plane and the lepton plane in the hadron frame.  Then the
lepton momentum can be parametrized as
\beq
k^\mu={Q\over 2}\left( \cosh\psi,\sinh\psi\cos\phi,
\sinh\psi\sin\phi,-1\right)
\label{eq2.lepton}
\eeq
and $k'^\mu=k^\mu-q^\mu$ with
\beq
\cosh\psi = {2x_{bj}S_{eA}\over Q^2} -1.
\label{eq2.cosh}
\eeq
With these definitions, the cross section for
(\ref{eq2.1}) can be expressed in terms of
$S_{ep}$, $x_{bj}$, $Q^2$, $z_f$, $q_T^2$ and $\phi$ in the hadron
frame.  Obviously, $\phi$ is invariant under boost in the $q$
direction, so that the $\phi$ in the hadron frame is the same,
for example, in the center-of-mass system of the virtual photon and the initial
proton $A$. 
The expression for these variables
in terms of the lab. frame variables is given in the Appendix.

\begin{figure}[ht]
\setlength{\unitlength}{1cm}
\begin{center}
\begin{picture}(10,19.5)
\psfig{file=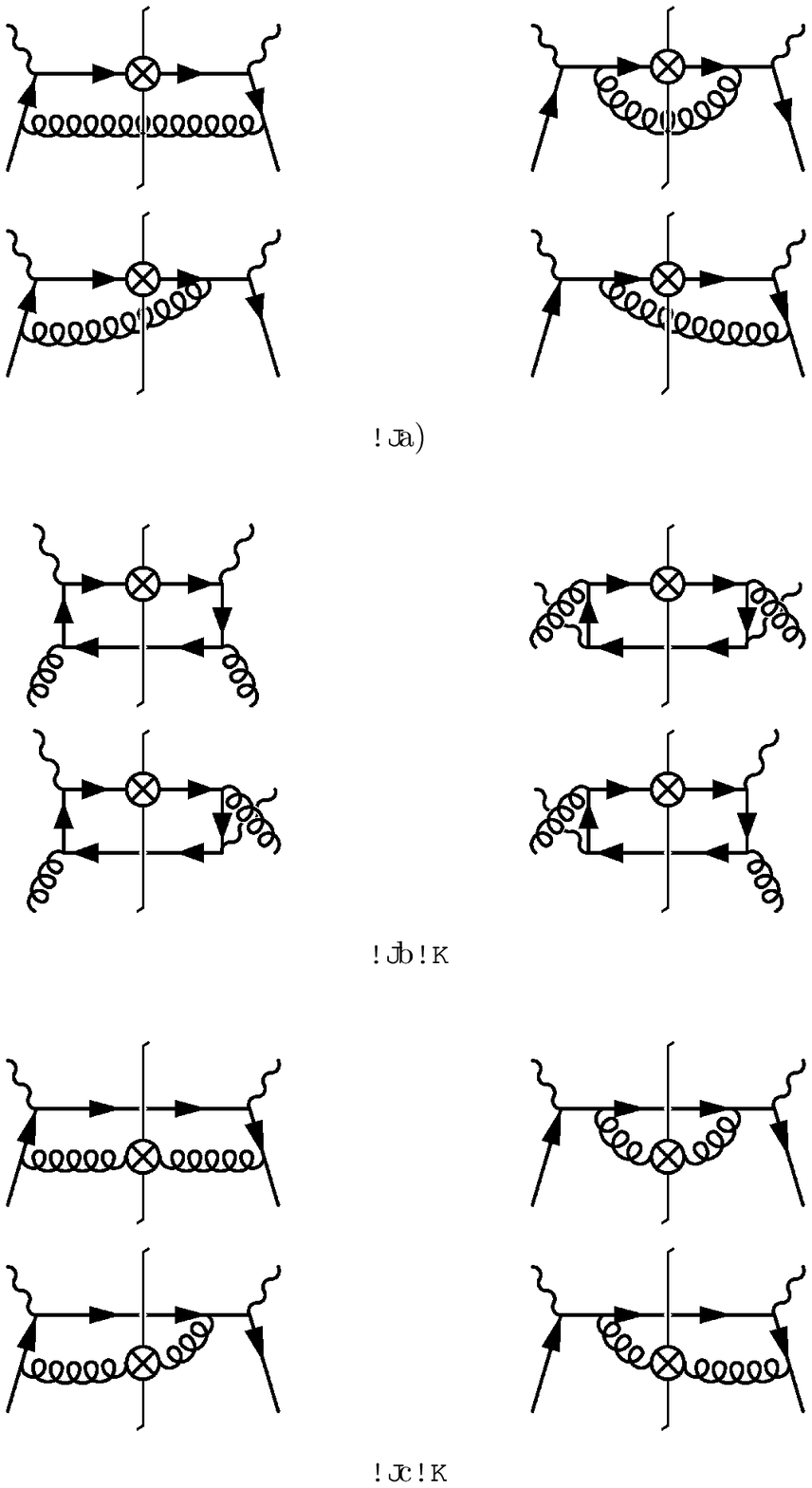,width=10cm}
\end{picture}
\caption{$O(\alpha_s)$ cut diagrams contributing to large $p_T$
hadron production in SIDIS.  Distribution 
function is located in the lower part of each diagram,
and $\otimes$ show the fragmentation 
function insertion.}
\end{center}
\end{figure}

Let us proceed to calculate the cross section.
As is shown in (\ref{eq2.p_B}) the transverse momentum of 
$p_B$ is $p_T=z_f q_T$ in the hadron frame, and we shall derive
the cross section formula differential in $q_T^2$.  
Lowest order diagram in the QCD coupling relevant for the 
processes (\ref{eq1.1}) is shown
in Fig. 1.
We follow the method of \cite{MOS92}
for the calculation.  To isolate the $\phi$-dependence in the cross section,
we first 
expand the hadronic tensor $W^{\mu\nu}$ in terms of the complete set of
the independent tensors.
To this end we
introduce the following 4 vectors which are orthogonal to
each other\,\cite{MOS92}:   
\beq
T^\mu &=&{1\over Q}\left(q^\mu + 2x_{bj}p_B^\mu\right),\nonumber\\
X^\mu &=&{1\over q_T}\left\{ {p_B^\mu \over z_f} -q^\mu -\left(
1+{q_T^2\over Q^2}\right)x_{bj} p_A^\mu\right\},\nonumber\\
Y^\mu &=& \epsilon^{\mu\nu\rho\sigma}Z_\nu X_\rho T_\sigma ,\nonumber\\
Z^\mu &=& -{q^\mu \over Q}.
\eeq
These vectors take the form of 
$T^\mu=(1,0,0,0)$, $X^\mu=(0,1,0,0)$, $Y^\mu=(0,0,1,0)$, $Z^\mu=(0,0,0,1)$
in the hadron frame, which greatly facilitates the actual calculation
presented below.  
Note that $T$, $X$ and $Z$ are vectors, while
$Y$ is an axial vector.
Since $W^{\mu\nu}$ satisfies the current conservation,
$q_\mu W^{\mu\nu} =q_\nu W^{\mu\nu}=0$, it is easy to see that
$W^{\mu\nu}$ can be expanded in terms of
9 independent tensors, for which we employ the following:
\beq
{\cal V}_1^{\mu\nu}&=&X^\mu X^\nu + Y^\mu Y^\nu,\quad
{\cal V}_2^{\mu\nu}=g^{\mu\nu} + Z^\mu Z^\nu,\quad
{\cal V}_3^{\mu\nu}=T^\mu X^\nu + X^\mu T^\nu,\nonumber\\
{\cal V}_4^{\mu\nu}&=&X^\mu X^\nu - Y^\mu Y^\nu,\quad
{\cal V}_5^{\mu\nu}= i(T^\mu X^\nu - X^\mu T^\nu),\quad
{\cal V}_6^{\mu\nu}=i(X^\mu Y^\nu - Y^\mu X^\nu),\nonumber\\
{\cal V}_7^{\mu\nu}&=&i(T^\mu Y^\nu - Y^\mu T^\nu),\quad
{\cal V}_8^{\mu\nu}=T^\mu Y^\nu + Y^\mu T^\nu,\quad
{\cal V}_9^{\mu\nu}=X^\mu Y^\nu + Y^\mu X^\nu.\nonumber\\
\eeq
The expansion coefficients of $W^{\mu\nu}$ in terms of these tensors
is easily obtained by the inverse tensors $\widetilde{{\cal V}}_k$  
for ${\cal V}_k$ ($k=1,\cdots 9$) as
\beq
{W}^{\mu\nu} =\sum_{k=1}^9 {\cal V}_k^{\mu\nu} \left[W_{\rho\sigma}
\widetilde{{\cal V}}_k^{\rho\sigma}\right],
\label{eq2.expansion}
\eeq
where
\beq
\widetilde{{\cal V}}_1^{\mu\nu}&=&{1\over 2}(2T^\mu T^\nu
+X^\mu X^\nu + Y^\mu Y^\nu),\quad
\widetilde{{\cal V}}_2^{\mu\nu}=T^\mu T^\nu,\quad
\widetilde{{\cal V}}_3^{\mu\nu}=-{1\over 2}(T^\mu X^\nu + X^\mu T^\nu),
\nonumber\\
\widetilde{{\cal V}}_4^{\mu\nu}&=&{1\over 2}(X^\mu X^\nu - Y^\mu Y^\nu),\quad
\widetilde{{\cal V}}_5^{\mu\nu} ={i\over 2}(T^\mu X^\nu - X^\mu T^\nu),\quad
\widetilde{{\cal V}}_6^{\mu\nu}={-i\over 2}(X^\mu Y^\nu - Y^\mu X^\nu),
\nonumber\\
\widetilde{{\cal V}}_7^{\mu\nu}&=&{i\over 2}(T^\mu Y^\nu - Y^\mu T^\nu),\quad
\widetilde{{\cal V}}_8^{\mu\nu}=-{1\over 2}(T^\mu Y^\nu + Y^\mu T^\nu),\quad
\widetilde{{\cal V}}_9^{\mu\nu}={1\over 2}(X^\mu Y^\nu + Y^\mu X^\nu).
\nonumber\\
\eeq
Note that ${\cal V}_k^{\mu\nu}$ ($k=1,\cdots,4,8,9$) are symmetric
between $\mu$ and $\nu$, and the rest are anti-symmetric.
${\cal V}_k^{\mu\nu}$ ($k=1,\cdots,5$) are even under parity,
while the rest are odd under parity.
For the processes with unpolarized electrons, i.e., (i)-(iii) in (\ref{eq1.1}),
both $L^{\mu\nu}$ and $W^{\mu\nu}$ are symmetric tensors, accordingly
${\cal V}^{\mu\nu}_k$ ($k=1,\cdots,4$) contributes to $W^{\mu\nu}$.  
For (iv) and (v)
both $L^{\mu\nu}$ and $W^{\mu\nu}$ are anti-symmetric pseudo-tensors
due to a $\gamma_5$ in the trace, 
accordingly only 
${\cal V}^{\mu\nu}_k$ ($k=6,7$) contribute to $W^{\mu\nu}$.  
Actual calculation of $\left[W_{\rho\sigma}
\widetilde{{\cal V}}_k^{\rho\sigma}\right]$ in (\ref{eq2.expansion})
involves trace over many $\gamma$-matrices,
which can be carried out conveniently in the hadron frame 
by using the Tracer program.

With these ${\cal V}^{\mu\nu}_k$, we can decompose the $\phi$-dependence of 
the cross sections by the contraction with $L^{\mu\nu}$ in (\ref{eq2.3}).
Define ${\cal A}_k$ ($k=1,\cdots,9$) as
\beq
{\cal A}_k={1\over Q^2}L_{\mu\nu}{\cal V}_k^{\mu\nu}
\eeq
then one obtains 
\beq
{\cal A}_1&=&1+\cosh^2\psi,\nonumber\\
{\cal A}_2&=&-2,\nonumber\\
{\cal A}_3&=&-\cos\phi\sinh 2\psi,\nonumber\\
{\cal A}_4&=&\cos 2\phi\sinh^2\psi,\nonumber\\
{\cal A}_6&=&-2\cosh\psi,\nonumber\\
{\cal A}_7&=&2\cos\phi\sinh\psi.
\label{eq2.Ak}
\eeq
From (\ref{eq2.Ak}), the cross section for
(i),(ii) and (iii) have $\phi$-dependence of
1, $\cos\phi$ and $\cos 2\phi$, while (iv) and (v) have only
1 and $\cos\phi$ dependence.

For the case of (iii) $e+p^\uparrow \to e'+ \Lambda^\uparrow +X$,
we need to parametrize the spin vector $S_{A\perp}$ and
$S_{B\perp}$ which lie in the plane orthogonal
to $\pvec_A$ and $\pvec_B$, respectively.   
We write
\beq
S_{A\perp}^\mu
&=&\cos\Phi_A X^\mu + \sin\Phi_A Y^\mu
=(0,\cos\Phi_A, \sin\Phi_A,0),
\label{eq2.spinA}\\
S_{B\perp}^\mu
&=&\cos\Theta_B\cos\Phi_B X^\mu + \sin\Phi_B Y^\mu
-\sin\Theta_B\cos\Phi_B Z^\mu\nonumber\\
&=&(0,\cos\Theta_B\cos\Phi_B, \sin\Phi_B,
-\sin\Theta_B\cos\Phi_B )  
\label{eq2.spinB}
\eeq
where the second equality in each equation follows 
only in the hadron frame.
In the hadron frame,
$\Phi_{A,B}$ are the azymuthal angles of 
$S_{A,B\perp}$ measured from the hadron plane, 
and $\Theta_B$ is the
polar angle of $\pvec_B$ measured from the $Z$-axis 
which has the Lorentz
invariant expression
\beq
\cos\Theta_B={q_T^2-Q^2\over q_T^2 + Q^2},\qquad
\sin\Theta_B={2q_TQ\over q_T^2 + Q^2}.
\eeq
In the Appendix, we present
the expression for $\Phi_A$ and $\Phi_B$ in terms of 
lab. frame variables.

\section{Analytic formula for cross section}
\setcounter{equation}{0}

By applying the method described in the previous section, we finally obtain
the cross section for the large $p_T$ hadron production in DIS 
in the following form:  
\beq
{d^5\sigma\over dx_{bj} dQ^2 dz_f dq_T^2 d\phi}
&=&{\alpha_e^2 \alpha_s \over 8\pi x_{bj}^2 S_{ep}^2 Q^2}
\sum_k {\cal A}_k \int_{x_{min}}^1\,{dx\over x}\int_{z_f}^1\,{dz\over z}\,
\left[ f\otimes D\otimes\widehat{\sigma}_k\right] \nonumber\\
& &\qquad\qquad\qquad\times\delta\left( {q_T^2\over Q^2} -
\left( {1\over \xhat} -1\right)\left({1\over \zhat}-1\right)\right),
\label{eq3.1}
\eeq
where ${\cal A}_k$ is defined in (\ref{eq2.Ak}), 
$\alpha_e=e^2/4\pi$ is the QED coupling constant, and 
we introduced the variables
\beq
\xhat={x_{bj}\over x},\qquad \zhat={z_f\over z},
\eeq
and 
\beq
x_{min}=x_{bj}\left( 1 + {z_f\over 1-z_f}{q_T^2\over Q^2}\right).
\label{eq3.2}
\eeq
For a given $S_{ep}$, $Q^2$ and $q_T$,
the kinematic constraint for $x_{bj}$ and $z_f$ is
\beq
& &{Q^2\over S_{ep}}<x_{bj}<1,
\label{range_zb}\\
& &0< z_f < {1-x_{bj}\over 
1-x_{bj}+x_{bj}q_T^2/Q^2}.
\label{range_zf}
\eeq  
For a given $Q^2$, $x_{bj}$ and $z_f$, $q_T$ can take
\beq
0< q_T < Q\sqrt{\left({1\over x_{bj}}-1\right)\left({1\over z_{f}}-1\right)}.
\eeq
Roughly speaking, 
$0< q_T <Q$ corresponds to the current fragmentation region and
$q_T> Q$ corresponds to the target fragmentation region.
As is shown in (\ref{eq2.p_B}), the 
transverse momentum of $p_B^\mu$ in the hadron frame is
\beq
p_{T}=z_f q_T <
z_fQ\sqrt{\left({1\over x_{bj}}-1\right)\left({1\over z_{f}}-1\right)}. 
\eeq

In (\ref{eq3.1}),
$\left[ f\otimes D\otimes\widehat{\sigma}_k\right]$ should read the
following form for each process in (\ref{eq1.1}) ($C_F=4/3$):

\vspace{0.5cm}

\noindent
(i) $e+p \rightarrow e' + \{\pi,\Lambda\} + X$\,\cite{Men}:\\
\beq
\left[f\otimes D\otimes \widehat{\sigma}_k\right]
&=&\sum_{q}e_q^2 q(x)\widehat{q}(z)\widehat{\sigma}_k^{qq}
+\sum_{q}e_q^2 G(x)\widehat{q}(z)\widehat{\sigma}_k^{gq}\nonumber\\
& &\qquad\qquad+\sum_{q}e_q^2 q(x)\widehat{G}(z)\widehat{\sigma}_k^{qg},
\eeq
where
\beq
\widehat{\sigma}_1^{qq}&=&2C_F\xhat\zhat\left\{{1\over Q^2q_T^2}
\left({Q^4\over \xhat^2\zhat^2} + \left(Q^2-q_T^2\right)^2\right) +6\right\},
\nonumber\\
\widehat{\sigma}_2^{qq}&=&2\widehat{\sigma}_4^{qq}=8C_F\xhat\zhat,\nonumber\\
\widehat{\sigma}_3^{qq}&=&4C_F\xhat\zhat{1\over Qq_T}(Q^2+q_T^2),
\label{eq3.avqq}
\eeq
\beq
\widehat{\sigma}_1^{gq}&=&\xhat(1-\xhat)\left\{{Q^2\over q_T^2}
\left({1\over \xhat^2\zhat^2} -{2\over \xhat\zhat}
+2\right) +10 -{2\over \xhat}-{2\over \zhat}\right\},
\nonumber\\
\widehat{\sigma}_2^{gq}&=&2\widehat{\sigma}_4^{gq}=8\xhat(1-\xhat),\nonumber\\
\widehat{\sigma}_3^{gq}&=&\xhat(1-\xhat){2\over Qq_T}\left\{2(Q^2+q_T^2)
-{Q^2\over \xhat\zhat}\right\},
\label{eq3.avgq}
\eeq
\beq
\widehat{\sigma}_1^{qg}&=&2C_F\xhat(1-\zhat)\left\{{1\over Q^2q_T^2}
\left({Q^4\over \xhat^2\zhat^2} + {(1-\zhat)^2\over \zhat^2}
\left(Q^2-{\zhat^2q_T^2\over (1-\zhat)^2}
\right)^2\right) +6\right\},
\nonumber\\
\widehat{\sigma}_2^{qg}&=&
2\widehat{\sigma}_4^{qg}=8C_F\xhat(1-\zhat),\nonumber\\
\widehat{\sigma}_3^{qg}&=&4C_F\xhat(1-\zhat)^2{1\over \zhat Qq_T}
\left\{Q^2+
{\zhat^2 q_T^2\over (1-\zhat)^2}\right\}.
\label{eq3.avqg}
\eeq

\noindent
(ii) $e+\vec{p} \rightarrow e' + \vec{\Lambda} + X$:\\
\beq
\left[f\otimes D\otimes \widehat{\sigma}_k\right]
&=&\sum_{q}e_q^2 \Delta q(x)\Delta\widehat{q}(z)\Delta_L\widehat{\sigma}_k^{qq}
+\sum_{q}e_q^2 \Delta G(x)\Delta\widehat{q}(z)\Delta_L\widehat{\sigma}_k^{gq}
\nonumber\\
& &\qquad\qquad+\sum_{q}e_q^2 \Delta q(x)\Delta\widehat{G}(z)
\Delta_L\widehat{\sigma}_k^{qg},
\eeq
where
\beq
\Delta_L\widehat{\sigma}_k^{qq}=\widehat{\sigma}_k^{qq}\qquad(k=1,2,3,4),
\eeq
\beq
\Delta_L\widehat{\sigma}_1^{gq}&=&
{(2\xhat-1)\left\{ Q^4 (\xhat-1)^2-q_T^4\xhat^2\right\}\over
Q^2 q_T^2 \xhat(\xhat-1)},
\nonumber\\
\Delta_L\widehat{\sigma}_2^{gq}&=&\Delta_L\widehat{\sigma}_4^{gq}=0,\nonumber\\
\Delta_L\widehat{\sigma}_3^{gq}&=&{2\left\{Q^2(\xhat-1)-q_T^2\xhat\right\}
\over Qq_T},
\eeq
\beq
\Delta_L\widehat{\sigma}_1^{qg}&=&2C_F\xhat\zhat\left\{
{\xhat-2\over \xhat-1} + {\xhat(\xhat+1)\over (\xhat-1)^2}
{q_T^4\over Q^4} +{2(2\xhat^2-2\xhat +1)\over (\xhat-1)^2}{q_T^2\over Q^2}
\right\},\nonumber\\
\Delta_L\widehat{\sigma}_2^{qg}&=&2\Delta_L\widehat{\sigma}_4^{qg}=
8C_F{\xhat^2\zhat\over \xhat-1}{q_T^2\over Q^2},\nonumber\\
\Delta_L\widehat{\sigma}_3^{qg}&=&{4C_F\xhat\zhat\over (\xhat-1)^2}
\left\{(\xhat-1)^2 + {\xhat^2q_T^2\over Q^2}\right\}{q_T\over Q}.
\eeq

\noindent
(iii) $e+{p}^\uparrow \rightarrow e' + {\Lambda}^\uparrow + X$:\\
\beq
\left[f\otimes D\otimes \widehat{\sigma}_k\right]
=\sum_{q}e_q^2 \delta q(x)\delta\widehat{q}(z)\Delta_T\widehat{\sigma}_k^{qq},
\eeq
where
\beq
\Delta_T\widehat{\sigma}_1^{qq}&=&4C_F{\rm cos}(\Phi_A-\Phi_B),\qquad
\Delta_T\widehat{\sigma}_2^{qq}=0,\nonumber\\
\Delta_T\widehat{\sigma}_3^{qq}&=&4C_F{Q\over q_T}{\rm cos}(\Phi_A-\Phi_B),
\qquad
\Delta_T\widehat{\sigma}_4^{qq}=4C_F{Q^2\over q_T^2}{\rm cos}(\Phi_A-\Phi_B).
\eeq

\noindent
(iv) $\vec{e}+\vec{p} \rightarrow e' + \{\pi,\Lambda\} + X$:\\
\beq
\left[f\otimes D\otimes \widehat{\sigma}_k\right]
&=&\sum_{q}e_q^2 \Delta q(x)\widehat{q}(z)\Delta_{LO}\widehat{\sigma}_k^{qq}
+\sum_{q}e_q^2 \Delta G(x)\widehat{q}(z)\Delta_{LO}\widehat{\sigma}_k^{gq}
\nonumber\\
& &\qquad\qquad
+\sum_{q}e_q^2 \Delta q(x)\widehat{G}(z)\Delta_{LO}\widehat{\sigma}_k^{qg},
\eeq
where
\beq
\Delta_{LO}\widehat{\sigma}_6^{qq}&=&-2C_F\left\{
\left({1\over \xhat\zhat} +\xhat\zhat\right){Q^2\over q_T^2}
-{\xhat\zhat q_T^2\over Q^2}\right\},\nonumber\\
\Delta_{LO}\widehat{\sigma}_7^{qq}&=&-4C_F\xhat\zhat{Q^2-q_T^2\over Q q_T},
\eeq
\beq
\Delta_{LO}\widehat{\sigma}_6^{gq}&=&-{2\xhat-1\over \xhat}
\left( 2\xhat + {\xhat-1\over \zhat^2}{Q^2\over q_T^2}\right),\nonumber\\
\Delta_{LO}\widehat{\sigma}_7^{gq}&=&-{2Q\over q_T}{
(\xhat-1)(2\zhat -1)\over \zhat},
\eeq
\beq
\Delta_{LO}\widehat{\sigma}_6^{qg}&=&{2C_F\zhat\over \xhat-1}\left\{
{1\over \zhat^2} - (\xhat-1)^2 + {\xhat^4\over(\xhat-1)^2}
{q_T^4\over Q^4}\right\},\nonumber\\
\Delta_{LO}\widehat{\sigma}_7^{qg}&=&{4C_F\xhat\zhat\over \xhat-1}
\left(1-{\xhat\over \zhat}\right){q_T\over Q}.
\eeq

\noindent
(v) $\vec{e}+{p} \rightarrow e' + \vec{\Lambda} + X$:\\
\beq
\left[f\otimes D\otimes \widehat{\sigma}_k\right]
&=&\sum_{q}e_q^2 q(x)\Delta\widehat{q}(z)\Delta_{OL}\widehat{\sigma}_k^{qq}
+\sum_{q}e_q^2 G(x)\Delta\widehat{q}(z)\Delta_{OL}\widehat{\sigma}_k^{gq}
\nonumber\\
& &\qquad\qquad
+\sum_{q}e_q^2 q(x)\Delta\widehat{G}(z)\Delta_{OL}\widehat{\sigma}_k^{qg},
\eeq
where
\beq
\Delta_{OL}\widehat{\sigma}_{6,7}^{qq}=\Delta_{LO}\widehat{\sigma}_{6,7}^{qq},
\eeq
\beq
\Delta_{OL}\widehat{\sigma}_6^{gq}&=&{2\xhat^2-2\xhat+1\over \xhat\zhat}
\left( \xhat + (\xhat-1){Q^2\over q_T^2}\right),\nonumber\\
\Delta_{OL}\widehat{\sigma}_7^{gq}&=&{2Q\over q_T}{
(\xhat-1)(2\xhat -1)\over \zhat},
\eeq
\beq
\Delta_{OL}\widehat{\sigma}_6^{qg}&=&{2C_F\zhat\over \xhat-1}\left\{
{1\over \zhat^2} + (\xhat-1)^2 - {\xhat^4\over(\xhat-1)^2}
{q_T^4\over Q^4}\right\},\nonumber\\
\Delta_{OL}\widehat{\sigma}_7^{qg}&=&-{4C_F\xhat\zhat\over \xhat-1}
\left(1-{\xhat\over \zhat}\right){q_T\over Q}.
\eeq

\vspace{0.5cm}

Several comments are in order here: 

\begin{enumerate}

\item[(1)]
From the above formula
one can separate the $\phi$-dependence 
of the $q_T$-differential cross section 
for (i)-(v) in (\ref{eq1.1}) as
\beq
{d\sigma\over dQ^2dx_{bj}dz_fdq_T^2 d\phi}
=\sigma_0 + {\rm cos}(\phi)\sigma_1 +
{\rm cos}(2\phi)\sigma_2,
\label{eq3.phi1}
\eeq
with $\sigma_2\equiv 0$ for (iv) and (v).
Transverse momentum of partons also gives the same $\phi$ dependence
as a purely kinematic effect\,\cite{Cahn}.
It is, however, interesting to test whether the above $0(\alpha_s)$ formula
gives quantitatively right magnitude for each component.

\item[(2)]
The cross section depends on $S_{ep}$ through ${\cal A}_k$ in
(\ref{eq2.Ak}).  For $x_{bj}S_{ep}/Q^2 \gg 1$ (see (\ref{eq2.cosh})), we have
${\cal A}_{1,3,4} \gg {\cal A}_{6,7}$ ($\phi$-dependence factored out), 
which results in the strong suppression of the asymmetry
for (iv) and (v) compared to (ii) and (iii) in (\ref{eq1.1}).

\item[(3)]
In the large-$q_T$ SIDIS,
contribution from the gluon distribution and fragmentation
functions is of the same $O(\alpha_s)$ effect 
as the quark contibution, so that $q_T$-differential 
cross section is expected to be a more sentitive tool
to determine detailed form of the polarized gluon contibution
than the $q_T$-integrated case\,\cite{MOS92}.

\item[(4)]
Except for (iii), $\sigma_0$ has a nonintegrable
$1/q_T^2$-dependence at $q_T\to 0$,
while $\sigma_{1,2}$ are integrable.
For the case of transverse polarization (iii), $\sigma_2$ has
the $1/q_T^2$-dependence, while $\sigma_{0,1}$
is integrable at $q_T\to 0$.  
To get a finite meaningful result for all value of $q_T$,
one needs to include higher order 
effect based on the resummation technique 
as in the case of the spin-average case\,\cite{MOS96,NSY00}.

\end{enumerate}

\section{Numerical estimate}
\setcounter{equation}{0}
\subsection{Azymuthal asymmetry}
In this section, we present a numerical estimate of the
polarized cross sections (ii)-(v) in (\ref{eq1.1})
in comparison with the spin averaged
one (i), using the exisiting parton distributions for the nucleon
and the fragmentation function for pion and $\Lambda$.
For this purpose we consider the azymuthal asymmetries
defined as follows:
\beq
\la 1\ra_{S_A S_B} &\equiv& { 
\int_0^{2\pi}\,d\phi {d^5\sigma^{pol} \over
dQ^2 dx_{bj} dz_f dq_T^2 d\phi} \over
\int_0^{2\pi}\,d\phi {d^5\sigma^{av} \over
dQ^2 dx_{bj} dz_f dq_T^2 d\phi}   }
={\sigma_0^{pol} \over \sigma_0^{av} },\nonumber\\
\la {\rm cos}(n\phi) \ra_{S_A S_B} &\equiv& { 
\int_0^{2\pi}\,d\phi\,{\rm cos}(n\phi) {d^5\sigma^{av,pol} \over
dQ^2 dx_{bj} dz_f dq_T^2 d\phi} \over
\int_0^{2\pi}\,d\phi {d^5\sigma^{av} \over
dQ^2 dx_{bj} dz_f dq_T^2 d\phi}   }
={\sigma_n^{av,pol} \over 2\sigma_0^{av} },\qquad (n=1,2)
\label{eq4.1}
\eeq
where $\sigma^{av}$
and $\sigma^{pol}$ are the spin-averaged and polarized
cross sections and their decomposition $\sigma_{0,1,2}^{av,pol}$ is 
defined in (\ref{eq3.phi1}). 
The subscript $S_A S_B$ in the left-hand-side of (\ref{eq4.1}) specify the
spin states of the hadrons $A$ and $B$
for each process in (\ref{eq1.1}).
We use the symbol $O$ for spin-average, 
$L$ for longitudinal polarization
and $T$ for transverse polarization.
Accordingly, $S_A S_B$ in (\ref{eq4.1}) can be 
$S_A S_B=OO,\ LL,\ TT,\ LO,\ OL$ corresponding to (i), 
(ii), (iii), (iv) and (v),
respectively.
By definition $\la 1\ra_{OO} \equiv 1$. 
For all cases, $|\la 1\ra_{S_A S_B}|<1$ and
$|\la {\rm cos}(n\phi) \ra_{S_A S_B}|<1$.  

The asymmetries (\ref{eq4.1}) are still functions of 
5 variables, $S_{ep}$, $Q^2$, $x_{bj}$, $z_f$ and $q_T$.
We estimate them 
at COMPASS ($S_{ep}=300$ GeV$^2$) and
EIC ($S_{ep}=10^4$ GeV$^2$) energies
with typical kinematic variables
where our perturbative formula are valid.
Comparison of predicted asymmetries with
experimental data may be better achieved by integrating
over some of the variables to gain statistics.  Here we will not try this
procedure, but will show the nonintegrated azymuthal asymmetry,  
intending to show a typical dependence of the asymmetry on
each variable.  
To determine the kinematic variables, 
we first note from (\ref{eq2.cosh}) and (\ref{eq2.Ak})
that at
large $\cosh\psi=2x_{bj}S_{ep}/Q^2-1$,
the spin asymmetry for (iv) and (v) is strongly suppressed 
compared with (ii) and (iii), so that a smaller value of $\cosh\psi$
is prefered to get large asymmetry.    
In addition,
if one chooses the same values of $Q^2$ and $\cosh\psi$ at COMPASS and
EIC energies, variation of the asymmetry at different $S_{ep}$ 
is wholly ascribed to 
the variation of distribution and/or fragmentation functions
and the partonic hard cross sections at different $x_{bj}$. 
Keeping this in mind we have chosen 
$(Q^2,x_{bj})=(100\ {\rm GeV}^2,0.4)$
for the COMPASS energy and $(Q^2,x_{bj})=(100\ {\rm GeV}^2,0.012)$ for the
EIC energy, 
by which ``valence'' and ``sea'' region of
the parton densities are mainly probed.

As a reference distribution and fragmentation functions, we use
GRV parton density for the unpolarized nucleon\,\cite{GRV98},
GRSV polarized parton density\,\cite{GRSV00}, KKP fragmentation
function for $\pi^+ + \pi^-$\,\cite{KKP} and FSV $\Lambda$-fragmentation 
function\,\cite{FSV98}
for polarized and unpolarized $\Lambda +\bar{\Lambda}$.  
In all cases we use NLO versions, 
putting the factorization scale $\mu_F^2=Q^2$.

\subsection{Pion or 2-jets production}
The process (iv) $\vec{e}+\vec{p}\to e'+{\rm 2\ jets}+X$ or $e'+\pi+X$ 
may be useful to get further insight into 
the polarized parton distribution in the nucleon. 
Here we present an estimate of
$\la 1\ra_{LO}$ and $\la\cos\phi\ra_{LO}$
for the charged pion production.
Figure 2 shows
the $q_T$-dependence ((a) and (b))
and $z_f$-dependence ((c) and (d)) of these asymmetries
at COMPASS ((a) and (c)) and EIC ((b) and (d)) kinematics.
$\la \cos\phi\ra_{OO}$ is also shown for comparison. 
For the spin asymmetry, we showed 
the results with standard (solid line) and 
valence (dashed line) scenarios in the GRSV distribution.
At $x_{bj}=0.4$, the spin asymmetry $\la 1 \ra_{LO}$ can be as large as
50 \%, while $\la\cos\phi\ra_{LO}$ is within a few \% level.
At $x_{bj}=0.012$, these magnitudes are somewhat reduced,
since the polarized parton density is more suppressed 
in the small $x$-region compared to the unpolarized parton density. 
At $x_{bj}=0.4$, the two scenarios
give totally different behavior for $\la 1\ra_{LO}$,
which can be, in principle, a sensitive 
test for the polarized parton density.
Typical magnitude of $\la \cos\phi\ra_{OO}$ is larger than 
$\la \cos\phi\ra_{LO}$ as expected, since the unpolarized parton 
density is larger than the polarized one.

\begin{figure}[ht]
\setlength{\unitlength}{1cm}
\begin{center}
\begin{picture}(14,15)
\psfig{file=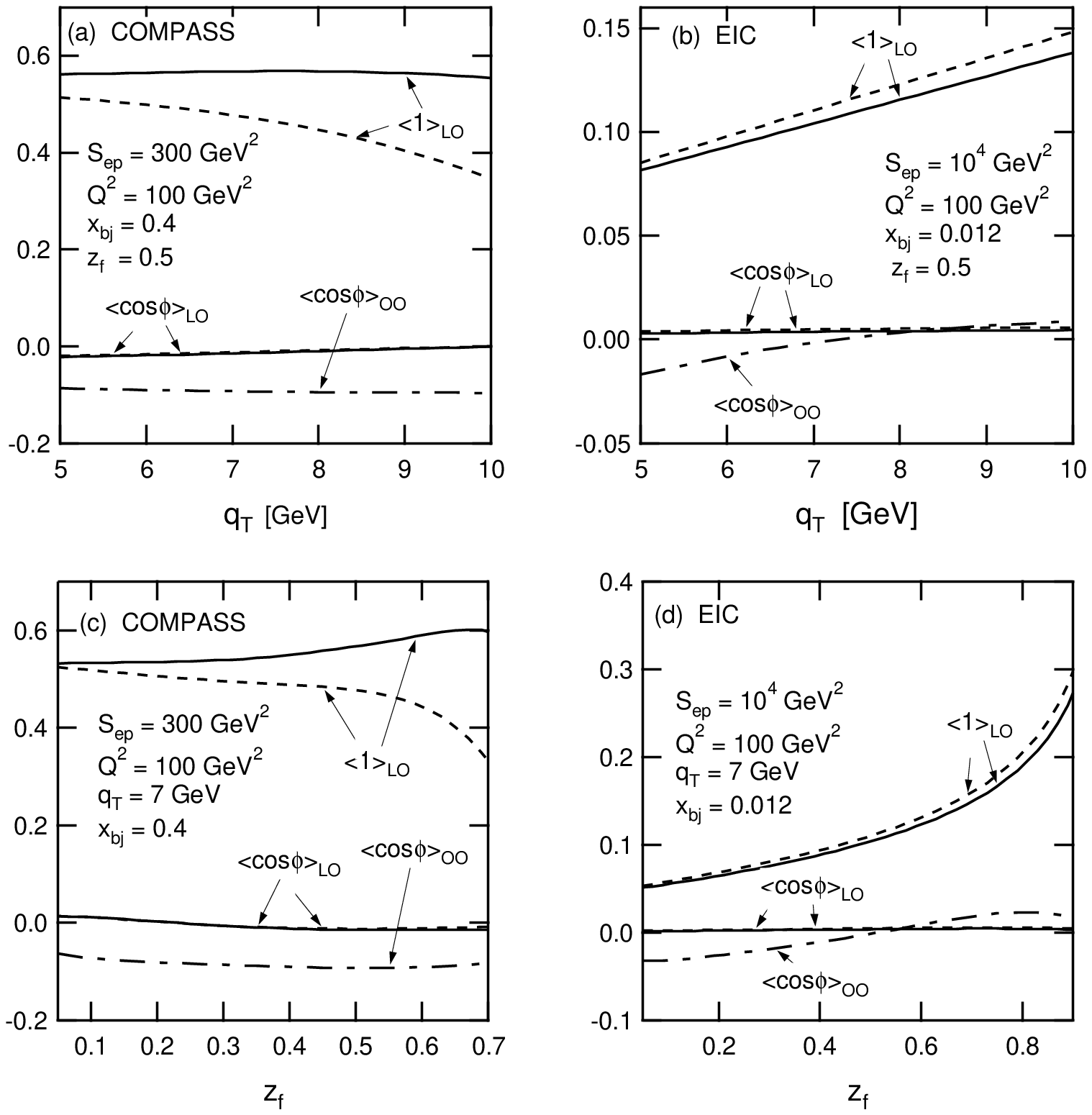,width=14cm}
\end{picture}
\caption{
Azymuthal spin asymmetry for the pion production
$\vec{e}+\vec{\pi}\to e'+ \pi +X$. (a) and (b) show the $q_T$-dependence
of the asymmetry at COMPASS and EIC energies, respectively.  
(c) and (d) show the $z_f$-dependence
of the asymmetry at COMPASS and EIC energies, respectively.
Solid- and dashed- lines, respectively, 
denote those obtained with standard- and valence- scenarios
of GRSV polarized parton density.  Dash-dot lines denotes
Azymuthal asymmetry $\la \cos\phi\ra_{OO}$ for the unpolarized
cross section.} 
\end{center}
\end{figure}

\subsection{$\Lambda$ production}

Figure 3 shows the 
$q_T$-dependence ((a) and (b)) and the $z_f$-dependence ((c) and (d))
of the azymuthal spin asymmetry for (v) $\vec{e}+p\to e'+\vec{\Lambda}+X$
((a) and (c) for $\la 1 \ra_{OL}$, and 
(b) and (d) for $\la \cos\phi\ra_{OL}$)
at the COMPASS kinematics.  Figure 4 shows the same asymmetries
as Fig. 3 but for the EIC kinematics.
In \cite{FSV98}, three sets of polarized 
fragmentation function for $\Lambda + \bar{\Lambda}$ 
have been constructed based on three
different assumptions
at a low energy input scale: 
Scenario 1 corresponds to the
nonrelativstic quark model
picture, where only polarized $s$-quark is assumed to fragment into
the polarized $\Lambda$, i.e., $\Delta\uhat=\Delta\dhat=0$.
Scenario 2 is based on the assumption that
the $\Lambda$ fragmentation function has a similar flavor decomposition as
the $\Lambda$ distribution function constructed from 
proton's structure function $g_1^p$
by the $SU(3)$ flavor symmetry,
i.e., $-\Delta\uhat=-\Delta\dhat=0.2\Delta\shat$. 
Scenario 3 assumes three flavors contribute equally to
the fragmentation process, i.e., $\Delta\uhat=\Delta\dhat=\Delta\shat$.
In all three cases, 
$\Delta\shat$ has positive sign. 
We calculated the asymmetries with these three sets. 
In both kinematics,
$\la 1 \ra_{OL}$ can be as large as 50 \%, while
$\la \cos\phi\ra_{OL}$ is within a few \% level,
varying with different
scenarios for the $\Lambda$ fragmentation functions.

\begin{figure}[ht]
\setlength{\unitlength}{1cm}
\begin{center}
\begin{picture}(14,15)
\psfig{file=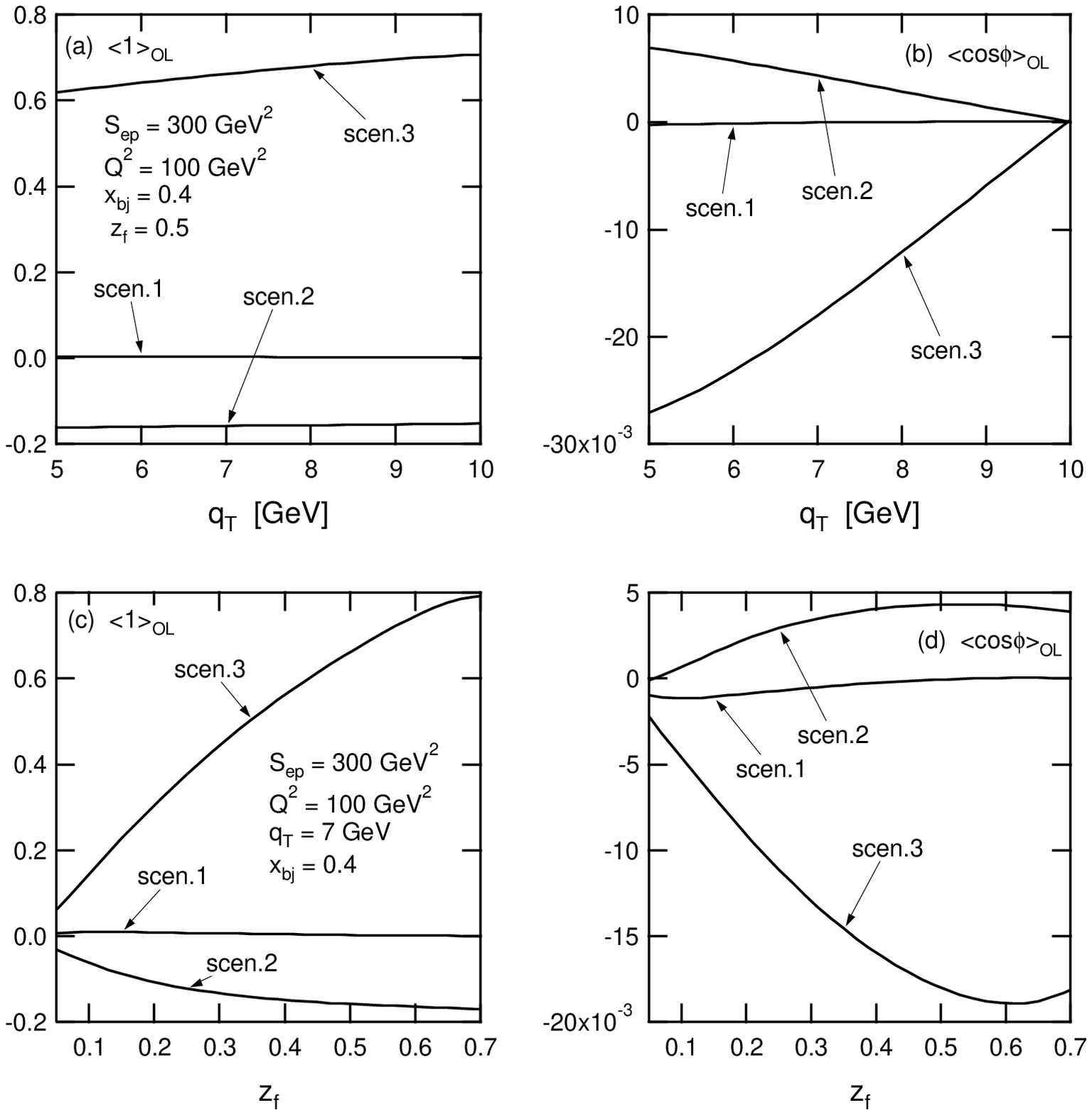,width=14cm}
\end{picture}
\caption{Azymuthal spin asymmetry for $\vec{e}+p\to e'+\vec{\Lambda}+X$
at the COMPASS energy.
(a) and (b) show the $q_T$-dependence of $\la 1\ra_{OL}$ and
$\la\cos\phi\ra_{OL}$, respectively. 
(c) and (d) show the $z_f$-dependence of $\la 1\ra_{OL}$
and $\la\cos\phi\ra_{OL}$, respectively. 
}
\end{center}
\end{figure}

The pattern of the calculated asymmetry
$\la 1 \ra_{OL}$ can be easily understood\,\cite{FSV98}.
At the COMPASS kinematics with $x_{bj}=0.4$,
contribution to the asymmetry is basically from the valence $u$-quark
distribution
(see (\ref{eq3.1}) and (\ref{eq3.2})).
Thus scenario 1 in which only $s$-quark fragments into $\vec{\Lambda}$
gives negligible asymmetry, while scenarios 2 and 3 gives, respectively, 
negative and positive asymmetries, the former being smaller in magnitude.
At the EIC kinematics with smaller $x_{bj}$ ($x_{bj}=0.012$),
a large contribution comes from the sea-quark distribution, which is
nearly flavor symmetric.    
Accordingly scenario 1 also gives a positive asymmetry
with its magnitude smaller than the scenario 3, since only $s$-quark 
fragments into $\vec{\Lambda}$.  In scenario 2, $u$ and $d$ quark contribution
is partially canceled by $s$-quark contribution, resulting in the 
small negative asymmetry.

\begin{figure}[ht]
\setlength{\unitlength}{1cm}
\begin{center}
\begin{picture}(14,15)
\psfig{file=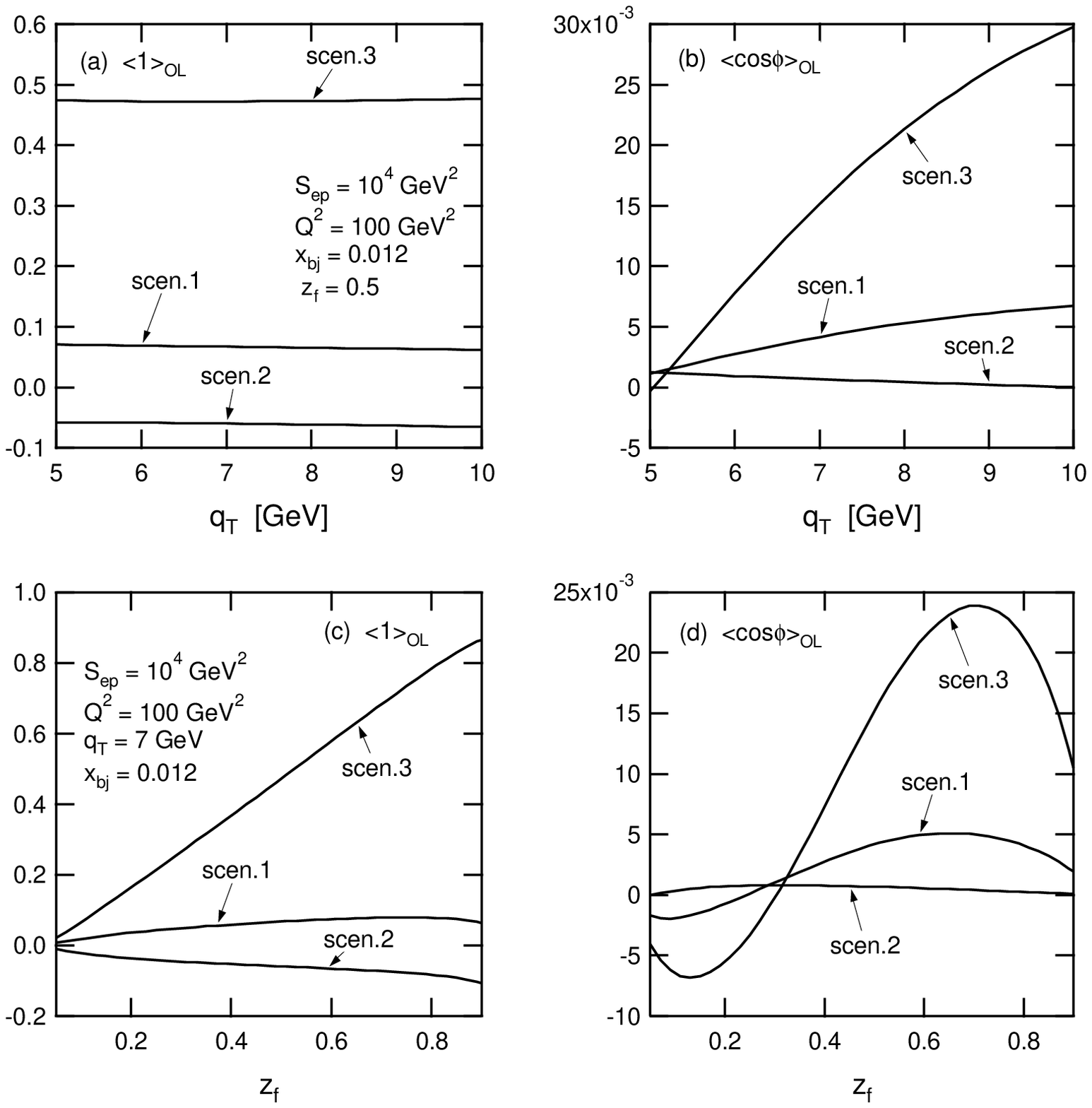,width=14cm}
\end{picture}
\caption{The same as Fig. 3 but for EIC energy.}
\end{center}
\end{figure}

Figures 5 and 6 show the asymmetry for 
(ii) $e+\vec{p} \to e'+\vec{\Lambda} +X$
at the COMPASS and EIC kinematics, respectively.
We show the results with the standard scenario 
for the GRSV polarized parton density. 
For comparison unpolarized azymuthal asymmetry
$\la\cos\phi\ra_{OO}$ and $\la\cos 2\phi\ra_{OO}$ are also shown
in Fig. 5, the former being essentially the same as the one for the 
pion production shown in Fig. 2 as expected. 
Typical magnitude of $\la 1\ra_{LL}$, $\la \cos\phi\ra_{LL}$ and
$\la \cos2\phi\ra_{LL}$ at $x_{bj}=0.4$ is
a few 10 \%, a few \%, and $O(10^{-3})$. 
At $x_{bj}=0.4$,
$\la 1\ra_{LL}$ and $\la \cos\phi\ra_{LL}$
have approximately the same magnitude as 
$\la 1\ra_{OL}$ and $\la \cos(\phi)\ra_{OL}$.  
At $x_{bj}=0.012$, $\la 1\ra_{LL}$ and $\la\cos\phi\ra_{LL}$
are, respectively, smaller than $\la 1\ra_{OL}$ and $\la\cos\phi\ra_{OL}$, 
since polarized parton density is more suppressed at smaller
$x$ region compared with unpolarized parton density. 
The distinction among three scenarios for the
polarized $\Lambda$ fragmentation function can be understood
in the same way
as the asymmetry for (v) $\vec{e}+p\to e'+\vec{\Lambda}+X$
shown in Figs. 3 and 4.

\begin{figure}[ht]
\setlength{\unitlength}{1cm}
\begin{center}
\begin{picture}(16.5,11)
\psfig{file=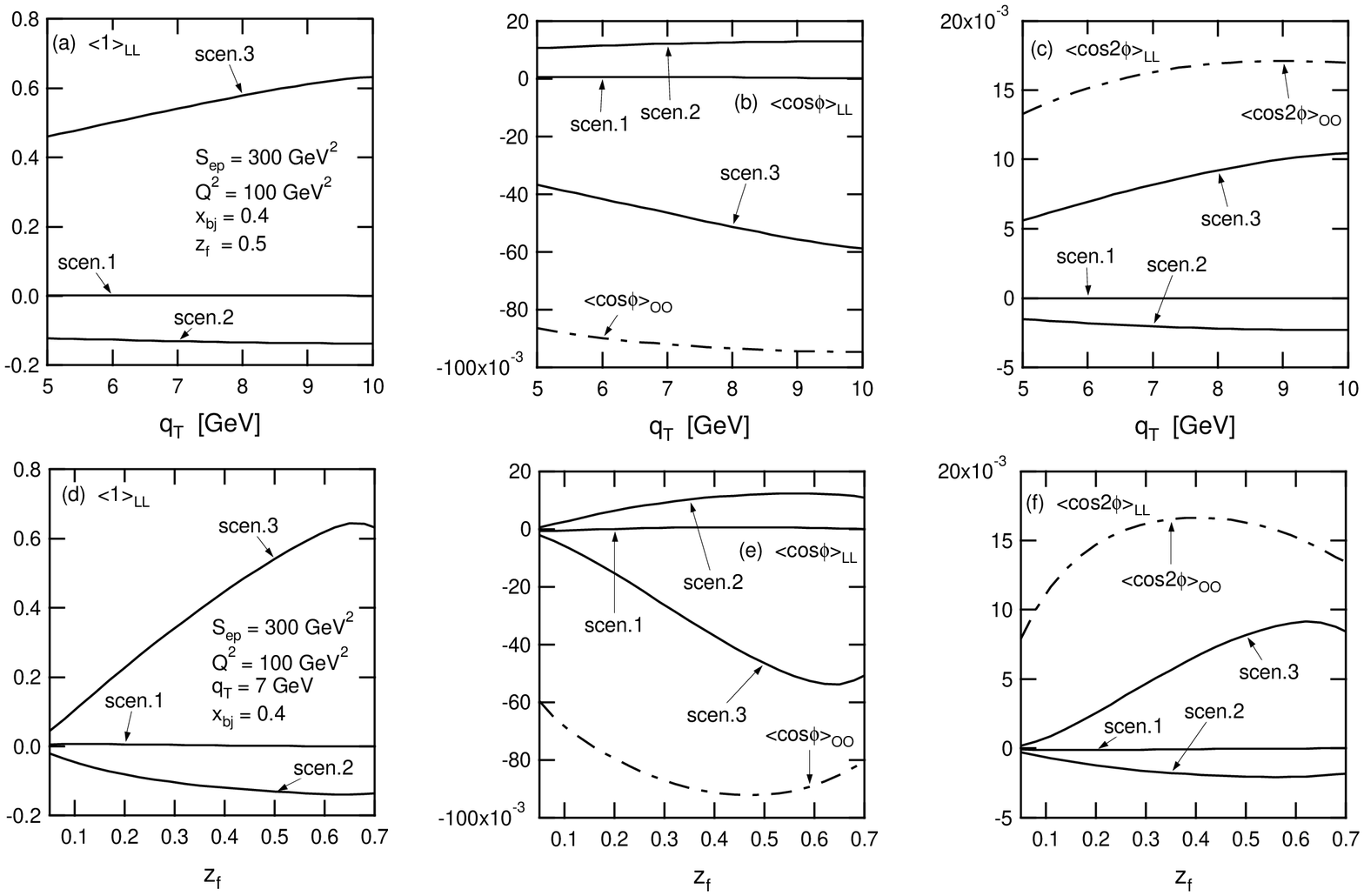,width=16.5cm}
\end{picture}
\caption{Azymuthal spin asymmetry for $e+\vec{p}\to e'+\vec{\Lambda}+X$
at the COMPASS energy.
(a),(b) and (c) show the $q_T$-dependence of $\la 1\ra_{LL}$,
$\la\cos\phi\ra_{LL}$ and $\la\cos2\phi\ra_{LL}$, respectively. 
(d),(e) and (f) show the $z_f$-dependence of $\la 1\ra_{LL}$,
$\la\cos\phi\ra_{LL}$ and $\la\cos2\phi\ra_{LL}$, respectively. 
Unpolarized azymuthal asymmetry $\la\cos\phi\ra_{OO}$
is shown in (b) and (e), and $\la\cos2\phi\ra_{OO}$ is shown in (c) and (f)
for comparison.}
\end{center}
\end{figure}

\begin{figure}[ht]
\setlength{\unitlength}{1cm}
\begin{center}
\begin{picture}(16.5,11)
\psfig{file=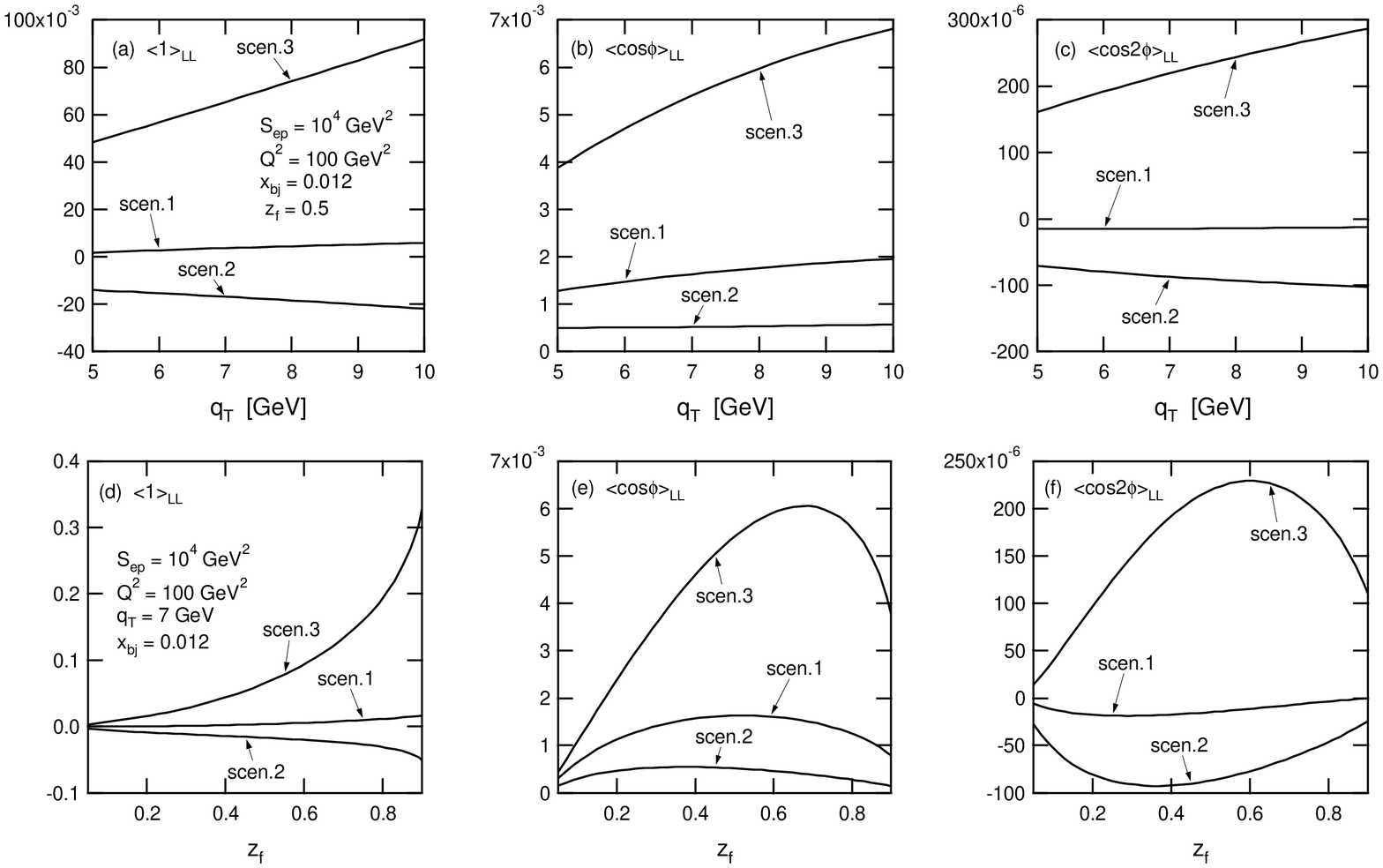,width=16.5cm}
\end{picture}
\caption{The same as Fig. 5 but for EIC energy. }
\end{center}
\end{figure}

Figures 7 and 8 show the spin asymmetry for
(iii) $e+{p}^\uparrow \to e'+{\Lambda}^\uparrow +X$ at
the COMPASS and EIC kinematics.
The calculation is done at
$\cos(\Phi_A-\Phi_B)=1$.  
In the present estimate,
we assume the 
transversity distribution and the trasversity fragmentation function
are equal to the longitudinally polarized distribution and
fragmentation function, respectively:
We put $\delta q \equiv \Delta q$ and $\delta \widehat{q} \equiv \Delta 
\widehat{q}$.   
This assumption may be justified at a low energy scale. 
But even in that case, they should be different at high energy,
since their $Q^2$-evolution is different, in particular, at 
small $x$\,\cite{AM}.  One can alternatively estimate the upper bound
based on the Soffer's equality as was done for
$p^\uparrow +p \to \Lambda^\uparrow +X$\,\cite{FSSV98}. 
Here we intend to present a simplest estimate 
to study characteristic features of the asymmetry by the above ansatz.  
One sees from the figures that
$\la 1\ra_{TT}$ is of a few \% $\sim 20$ \% 
at $x_{bj}=0.4$ depending on the
scenarios for the $\Lambda$ fragmentation function.   
This value is smaller than $\la 1\ra_{LL}$, 
which is expected since there is no gluon contribution to
the transverse spin asymmetry.
On the other hand,  
$\la \cos\phi\ra_{TT}$ is from a few to 10 \%, and  
even $\la \cos2\phi\ra_{TT}$ is of a few \% level, which are
much larger than 
$\la \cos n\phi\ra_{LL}$ ($n=1,2$). 
The same feature persists also at $x_{bj}=0.012$, although 
absolute magnitude of the asymmetry becomes smaller
by the same reason for $\la \cos n\phi \ra_{LL}$ ($n=0,1,2$).
It is interesting to check these peculiar features of the 
hard cross section
in future experiments
and to get information about transversity.

\begin{figure}[ht]
\setlength{\unitlength}{1cm}
\begin{center}
\begin{picture}(16.5,11)
\psfig{file=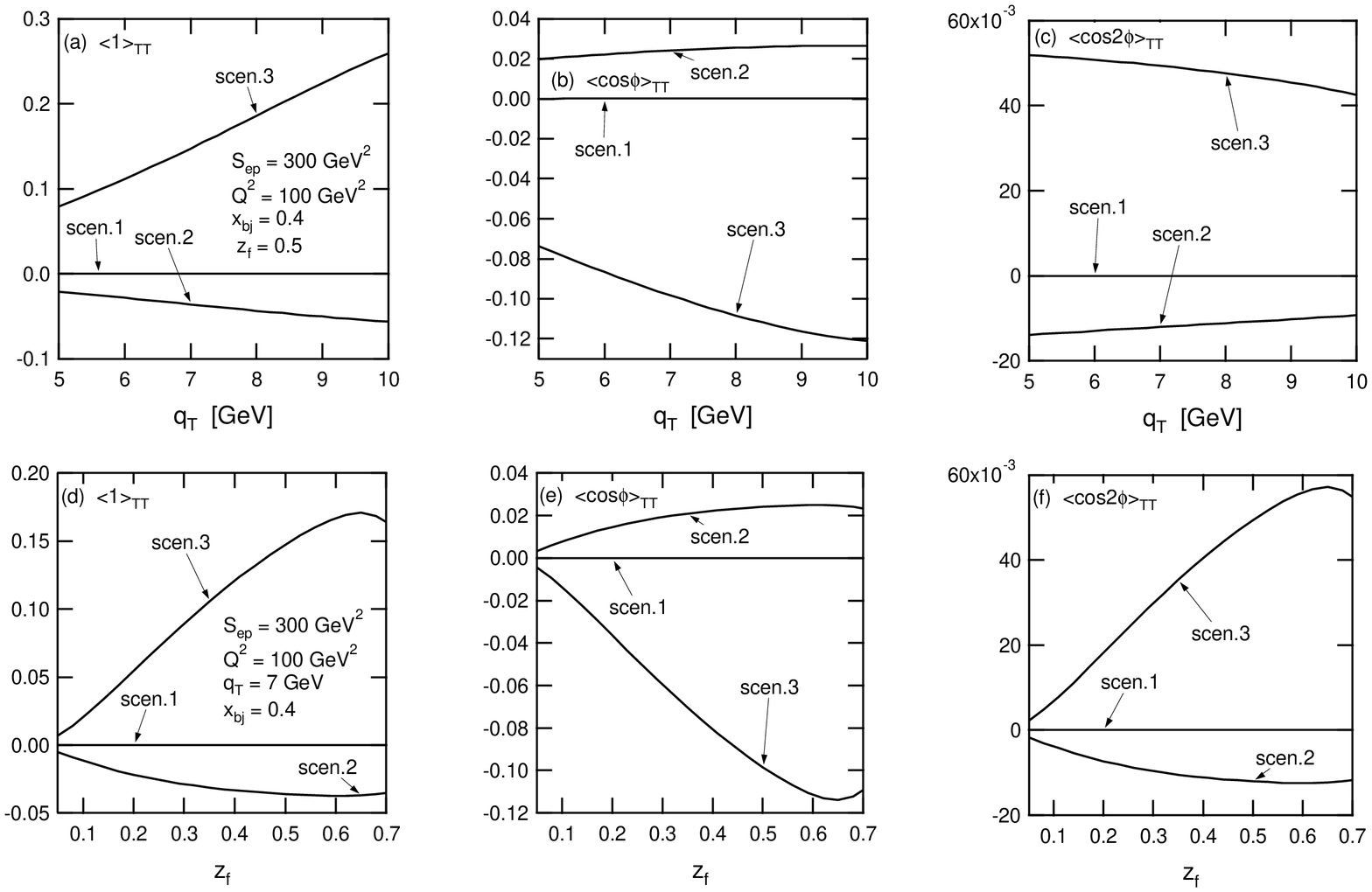,width=16.5cm}
\end{picture}
\caption{Azymuthal spin asymmetry for $e+{p}^\uparrow\to e'
+{\Lambda}^\uparrow+X$
at the COMPASS energy.
(a),(b) and (c) show the $q_T$-dependence of $\la 1\ra_{TT}$,
$\la\cos\phi\ra_{TT}$ and $\la\cos2\phi\ra_{TT}$, respectively.
(d),(e) and (f) show the $z_f$-dependence of $\la 1\ra_{TT}$,
$\la\cos\phi\ra_{TT}$ and $\la\cos2\phi\ra_{TT}$, respectively. 
}
\end{center}
\end{figure}

\begin{figure}[ht]
\setlength{\unitlength}{1cm}
\begin{center}
\begin{picture}(16.5,11)
\psfig{file=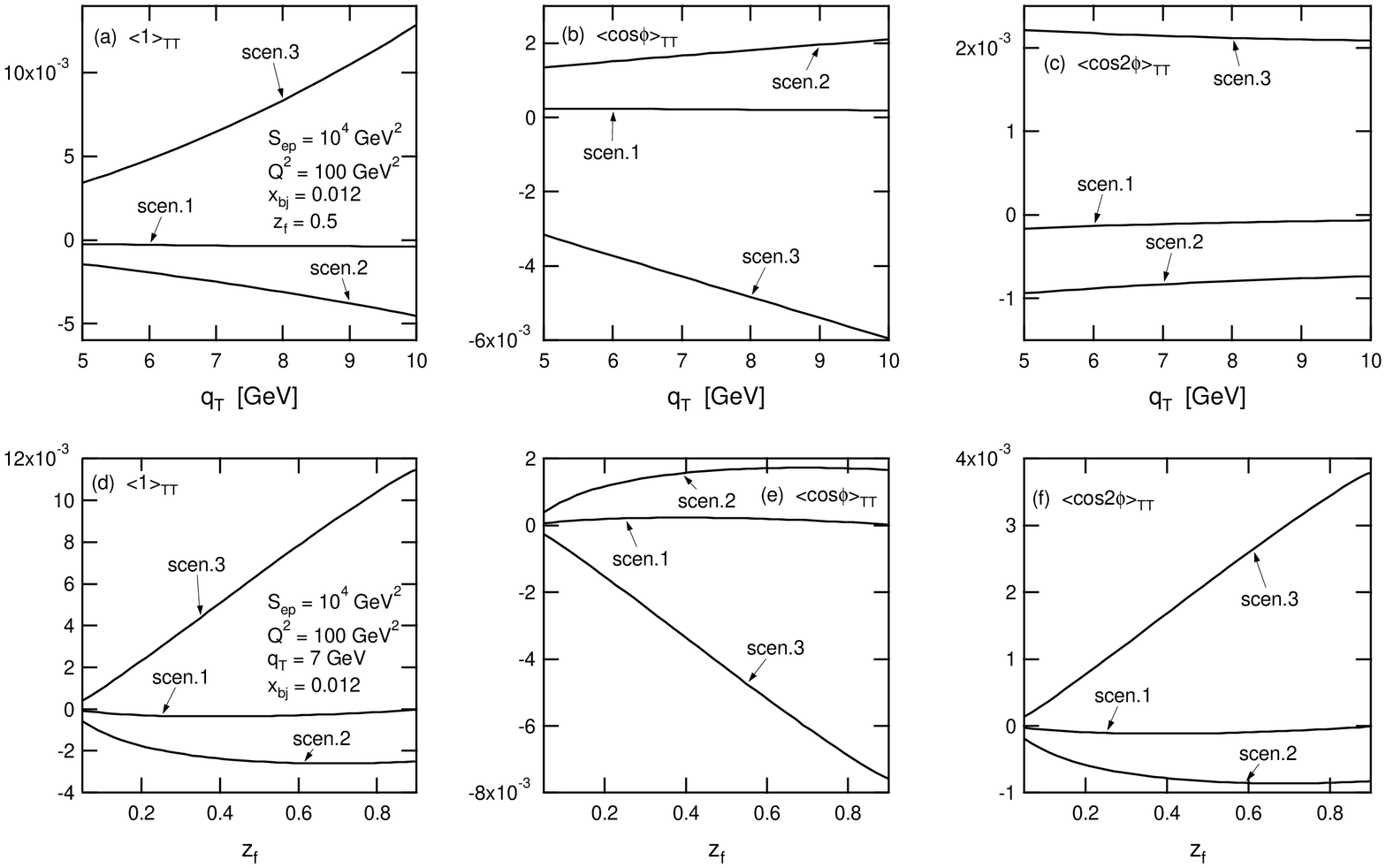,width=16.5cm}
\end{picture}
\caption{The same as Fig. 7 but for EIC energy.}
\end{center}
\end{figure}

\section{Summary and conclusion}
\setcounter{equation}{0}

In this paper, we have studied the double spin asymmeties
for 2-jets and large-$p_T$ hadron production in semi-inclusive DIS
with one-photon exchange.  
The cross section formula for the complete set of the spin dependent 
processes (\ref{eq1.1}) 
has been derived to $O(\alpha_s)$.  
Separating the dependence on the azymuthal angle $\phi$ between 
the hadron and lepton planes, the formula can be written as 
\beq
{d^5\sigma\over dx_{bj}dQ^2 dz_f dq_T^2 d\phi}
=\sigma_0 + \cos\phi\,\sigma_1 + \cos 2\phi\,\sigma_2,
\label{eq5.1}
\eeq
where the last term is absent for the process with polarized lepton
beams.  This decomposition is purely kinematic, 
while the magnitude of each component depends on
the origin of the
transverse momentum of the final hadron.  
At low $p_T$, nonperturbative component
such as intrinsic-$k_T$ of partons is expected to contribute
as studied in \cite{MT96}.  At moderate $p_T$, higher order effect
represented by the resummed cross section is presumably important. 
Our $O(\alpha_s)$ formula should be useful to clarify the importance of
these effects and to  
see at which $p_T$ the experimentally observed
spin asymmetry follows the perturbative QCD prediction. 

In order to see the qualitative behavior of each component in 
(\ref{eq5.1}), we calculated the azymuthal asymmetries
using the parton distribution for the nucleon and the fragmentation
functions for $\pi$ and $\Lambda$ 
at $x_{bj}=0.4$ and $x_{bj}=0.012$.  
We found that the asymmetry is very sensitive to 
different scenarios of parton distribution and fragmentation functions
as was expected from the previous studies on SIDIS and pp 
collisions\,\cite{FSV98,FSV98pp,FSSV98}.  
Typical magnitude of asymmetry from each component in (\ref{eq5.1})
turns out to be $O(0.1\sim 0.6)$, $O(10^{-2})$ and  $O(10^{-3})$
from the first to third term in (\ref{eq5.1}).  
For the process (iii) $e+p^\uparrow \to e' +\Lambda^\uparrow+X$
in (\ref{eq1.1}), however,
the second and third contribution in (\ref{eq5.1}) is significantly larger,
while the first term is smaller, compared with other spin asymmetries. 

As in the case of unpolarized cross sections, 
$\sigma_0$ component of the polarized cross sections
have $1/q_T^2$-singularity at low $q_T$, except for the above process
(iii) where $\sigma_2$ has the same singularity.  To get a meaningful
formula at low $q_T$, we need resummation technique\,\cite{MOS96,NSY00}, 
which will be 
reported in a future publication.

\vspace{0.7cm}
\large

\noindent
{\bf Acknowledgement:} \\[5pt]
\normalsize
This work is supported in part 
by the Grant-in-Aid for Scientific Research of Monbu-Kagaku-sho.  

\vspace{0.7cm}
\appendix
\setcounter{equation}{0}
\renewcommand{\theequation}{A.\arabic{equation}}
\large
\noindent
{\bf Appendix}\\[5pt]
\normalsize
\noindent
In this appendix, we give the expression
for the Lorentz invariants and hadron frame variables in terms of
the variables in the Lab. frame. (See \cite{MOS92} for the detail.)
In the Lab. frame, 
4-vectors which appeared in the text
can be parametrized as
\beq
p_A^\mu &=&E_A(1,0,0,1),\nonumber\\
p_B^\mu &=&E_B(1,\sin \theta_B \cos \phi_B,\sin \theta_B \sin \phi_B,
\cos \theta_B),\nonumber\\
k^\mu &=& E(1,0,0,-1),\nonumber\\
k'^\mu &=& E'(1,-\sin \theta,0, -\cos \theta ),\nonumber\\ 
q^\mu &=& (E-E',E'\sin \theta, 0, E'\cos \theta -E),\nonumber\\
S_{A\perp}^\mu &=& (0,\cos \Phi_A^L, \sin \Phi_A^L, 0),\nonumber\\
S_{B\perp}^\mu &=& (0,
\cos\theta_B\cos\phi_B\cos\Phi_B^L-\sin\phi_B\sin\Phi_B^L,\nonumber\\
& &\quad\cos\theta_B\sin\phi_B\cos\Phi_B^L+\cos\phi_B\sin\Phi_B^L,
-\sin\theta_B\cos\Phi_B^L),
\eeq
where $\Phi_{A,B}^L$ is the azymuthal angle of the transverse spin vector
$S_{A,B\perp}$ which satisfy $\vec{S}_{A\perp}\cdot \vec{p}_A =
\vec{S}_{B\perp}\cdot \vec{p}_B =0$.
In order to give the expression for $q_T^2$, we introduce
the polar angle $\theta_*$ for $p_B$ in the Born amplitude
(diagram obtained by removing the gluon line from Fig.1(a)),
which allows us to write
$p_B \propto (1, \sin\theta_*,0,\cos\theta_*)$.
It is easy to show $\theta_*$ is given as
\beq
\cot{\theta_*\over 2} = {2x_{bj}E_A\over Q}\sqrt{1-{Q^2\over x_{bj}S_{ep}}}.
\eeq
With this $\theta_*$ we have 
\beq
q_T^2 = {8E^2-4E'(2E-E')(1+\cos\theta)\over 1-\cos\theta_B}
\left\{\sin^2\left((\theta_B-\theta_*)/ 2\right)+
\sin\theta_B\sin\theta_*\sin^2\left(\phi_B/2\right)\right\},
\eeq
and
\beq
\cos\phi={Q\over 2q_T}\left(1-{Q^2\over x_{bj}S_{ep}}\right)^{-1/2}
\left\{1-{Q^2\over x_{bj}S_{ep}}+{q_T^2\over Q^2}-\left({Q\over 2x_{bj}E_A}
\right)^2\cot^2(\theta_B/2)\right\}.
\eeq
Finally, the azymuthal angles
$\cos\Phi_A$ and $\cos\Phi_B$ in the hadron frame can be written as
\beq
\cos\Phi_A &=& -S_{A\perp}\cdot X = 
{1\over q_T}\left( {1\over z_f}E_B\sin\theta_B\cos(\phi_B -\Phi_A^L)
-E'\sin\theta\cos\Phi_A^L\right),\\
\cos\Phi_B &=& {1\over \sin\Theta_B}S_{B\perp}\cdot Z\nonumber\\
&=&{Q^2+q_T^2\over 2Q^2q_T}\left\{ E'\sin\theta\left(
\cos\phi_B\cos\theta_B\cos\Phi_B^L - \sin\phi_B\sin\Phi_B\right)
\right.\nonumber\\
& &\left.\qquad\qquad-(E'\cos\theta -E)\sin\theta_B\cos\Phi_B^L \right\}.
\eeq

\end{document}